\documentclass[12pt,a4paper]{article}
\usepackage[dvips]{graphicx}
\begin{document}

\title{ 
\vspace{-4cm}
  \begin{flushright}
    \begin{small}
      {DPNU-00-25 \\ August, 2000 }
    \end{small}
  \end{flushright}
\vspace{2cm}
  Emergence of power-law correlation in 
  1-dimensional self-gravitating system
}

\author{
  Hiroko KOYAMA\thanks{\tt hiroko@allegro.phys.nagoya-u.ac.jp} \, 
and Tetsuro KONISHI\thanks{\tt tkonishi@allegro.phys.nagoya-u.ac.jp} \\
Department of Physics, School of Science,\\
Nagoya University, 464-8602, Nagoya, Japan
}
\date{}
\maketitle

\begin{abstract}
  A new phase of temporal evolution of the one-dimensional 
  self-gravitating system is numerically discovered.
  Fractal structure is dynamically created from non-fractal
  initial conditions.
  Implication to astrophysics and mathematical physics is discussed.
\end{abstract}

\baselineskip 24pt

\section{Introduction}
%

One-dimensional self-gravitating systems is a  simplified
model of 
gravitational many body system. 
The system represents the motion of parallel flat mass sheets  
interacting  each other through Newtonian gravity.
Since its equation of motion can be solved piecewise exactly
during each interval between collisions, numerical integration of the 
system is particularly simple. Hence the system is used to 
study long-time behavior and relaxation process 
of gravitational systems~\cite{hohl-feix-1967,
severne-1984-apss,miller-1996-pre,sheet-tgk-3}.
In addition,  
one-particle distribution for canonical and 
microcanonical ensemble 
is obtained~\cite{rybivki-1971-apss}.

\begin{figure}[hbtp]
  \begin{center}
    \includegraphics[width=7cm]{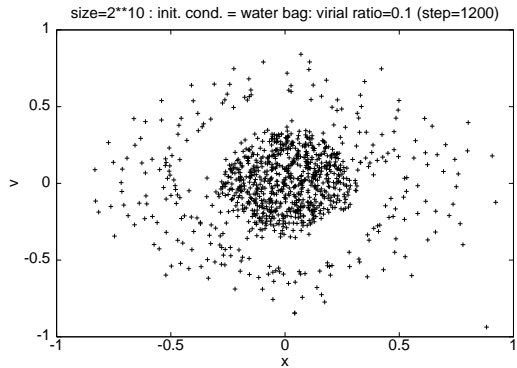}
    \caption{Particle distribution in $\mu$-space of the 
      model~(\ref{eq:sheet-hamiltonian}) for
      $ N=2^{10}$,$V_r = 0.1$. One big cluster is formed and 
       fractal structure is not present.
      E=0.25, Time=9.375.}
    \label{fig:ygs-1}
  \end{center}
\end{figure}

Yamashiro et al. studied this model numerically to investigate
relaxation process of elliptical 
galaxies~\cite{yamashiro-gouda-sakagami}.  According to the virial theorem
the virial ratio $\displaystyle V_r \equiv 2E_{kin}/E_{pot}$ for this 
model is expected to relax to 1.
They chose  initial conditions with $10^{-3} \le V_r \le 1$. 
Typical feature of the system they  described  is shown
in Fig.~\ref{fig:ygs-1},
where a snapshot  of distribution in $(x,u)$ space ($\mu$-space)
 is shown.
We can see a big cluster is formed. 

While their main concern was the process of relaxation,
they also noted that initial conditions with smaller
virial ratio ( typically  $V_r < 10^{-3}$ ) 
did not guarantee the relaxation within their CPU time.
They also observed formation of lumped structure as 
observed in \cite{hohl-feix-1967}, which are quite different 
from the one shown in Fig.~\ref{fig:ygs-1}. 
Anomaly in relaxation for  small value of  $V_r$ is also pointed out
in ~\cite{Miller-1}.

We extend their study to the region $V_r \rightarrow 0$ and found
that there is a new phase of evolution where a fractal
spatial structure  is spontaneously created.
We would like to stress that the fractal structure is created 
even when the initial condition is not fractal (typically 
uniformly random).  We also note that not all the initial conditions
lead to fractal structure.

In this letter we report the discovery of this novel phase.
In the next section we describe the  model.
We present the numerical results   in section 3, and the final
section is for summary and discussions.

\section{Model}
The one-dimensional self-gravitating system (also known as 
`mass-sheet model') consists of $N$ equivalent flat sheets of constant 
mass density. The sheets are infinitely extended  in the $y$ and $z$
direction and are aligned parallel to each other and move in
the  $x$ direction. 

If we suppose the sheets interact with Newtonian gravity,
the Hamiltonian of the system reads
\begin{eqnarray}
  H &=& \sum_{i=1}^N \frac{p_i^2}{2m} + 2\pi G m^2\sum_{i > j} \left|
x_i - x_j\right| \ , 
  \label{eq:sheet-hamiltonian} 
\end{eqnarray}
where $p_i = mu_i$. ($u_i$ is the velocity of the sheet $i$ \ .)
 We set $ m= 1/N$  so that the total mass is unity, 
 and $4\pi G \equiv 1 $.
 Time is measured in the unit 
\[
t_c \equiv 1/\sqrt{4\pi G M/L} 
\]
where $M \equiv mN$ is total mass, $L$ is the spatial length 
on which particles are distributed initially. 
$t_c$ is called `crossing time' and is a typical time for a particle
to traverse the system.

With this Hamiltonian, the equation of motion reads
\begin{equation}
  \ddot{x_i} = \frac{1}{2N}\left( N_R - N_L\right)
  \label{eq:eq-of-motion}
\end{equation}
where $N_R$ and $N_L$ are number of sheets which are on the right
and left of the sheet $i$ , respectively.

Eq.~(\ref{eq:eq-of-motion}) indicates that the force acting on each sheet
is constant until a collision  occurs between two sheets. 
Hence we can exactly 
integrate the equation of motion~(\ref{eq:eq-of-motion}) between
 every two collisions.
This fact greatly enhances the usefulness of this model.

\section{Numerical results : structure formation}

\begin{figure}[hbtp]
  \begin{center}
    \includegraphics[width=7cm]{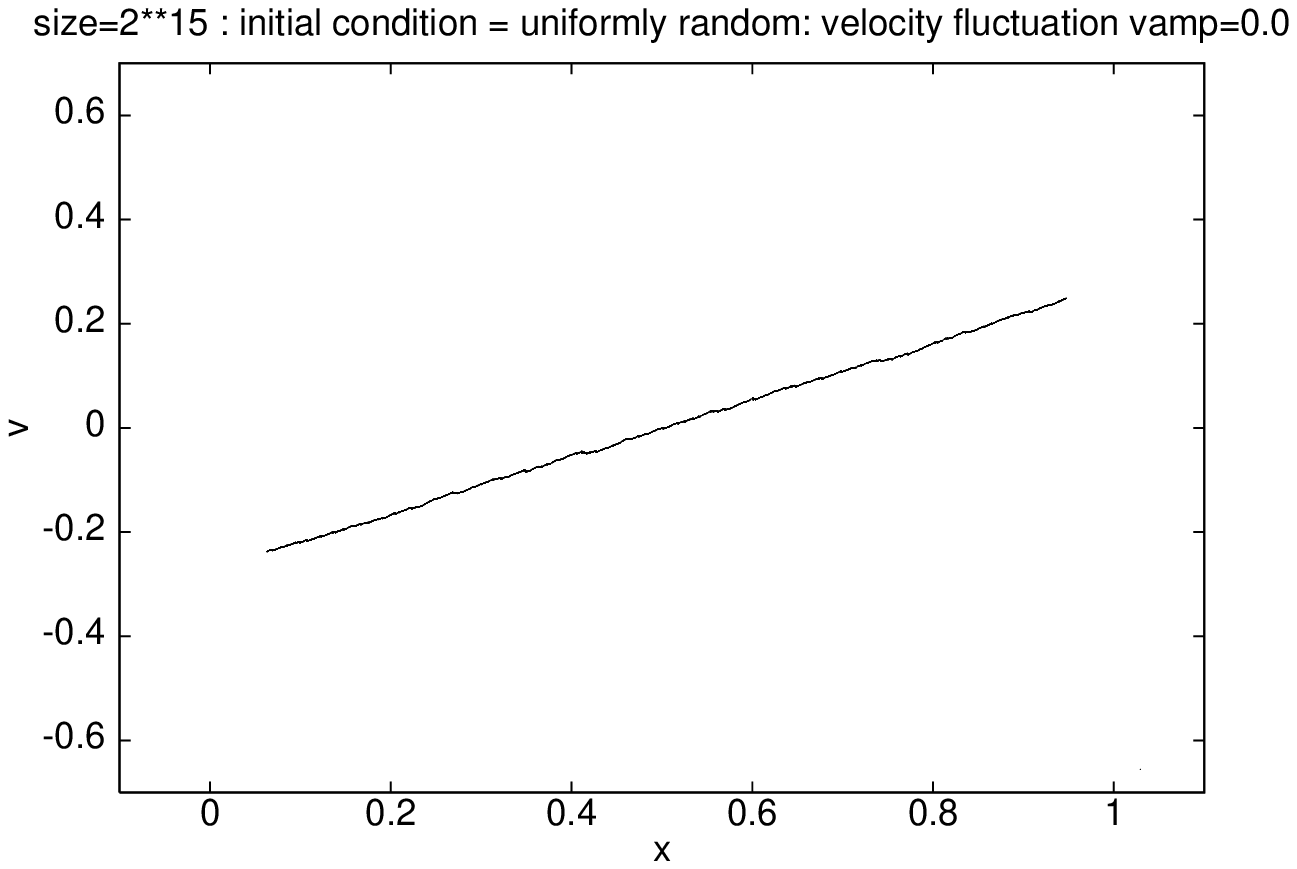}
    \includegraphics[width=7cm]{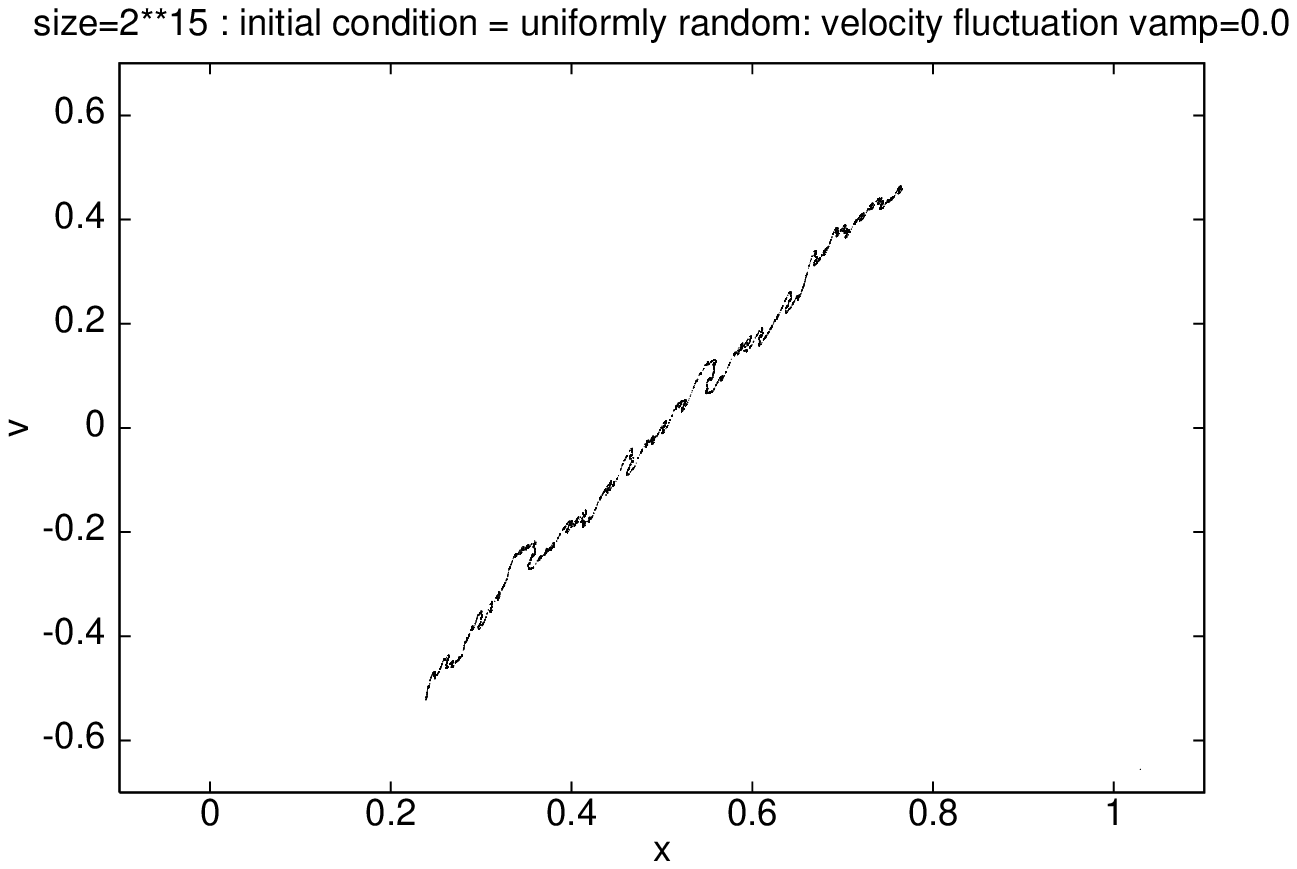}
    
    \includegraphics[width=7cm]{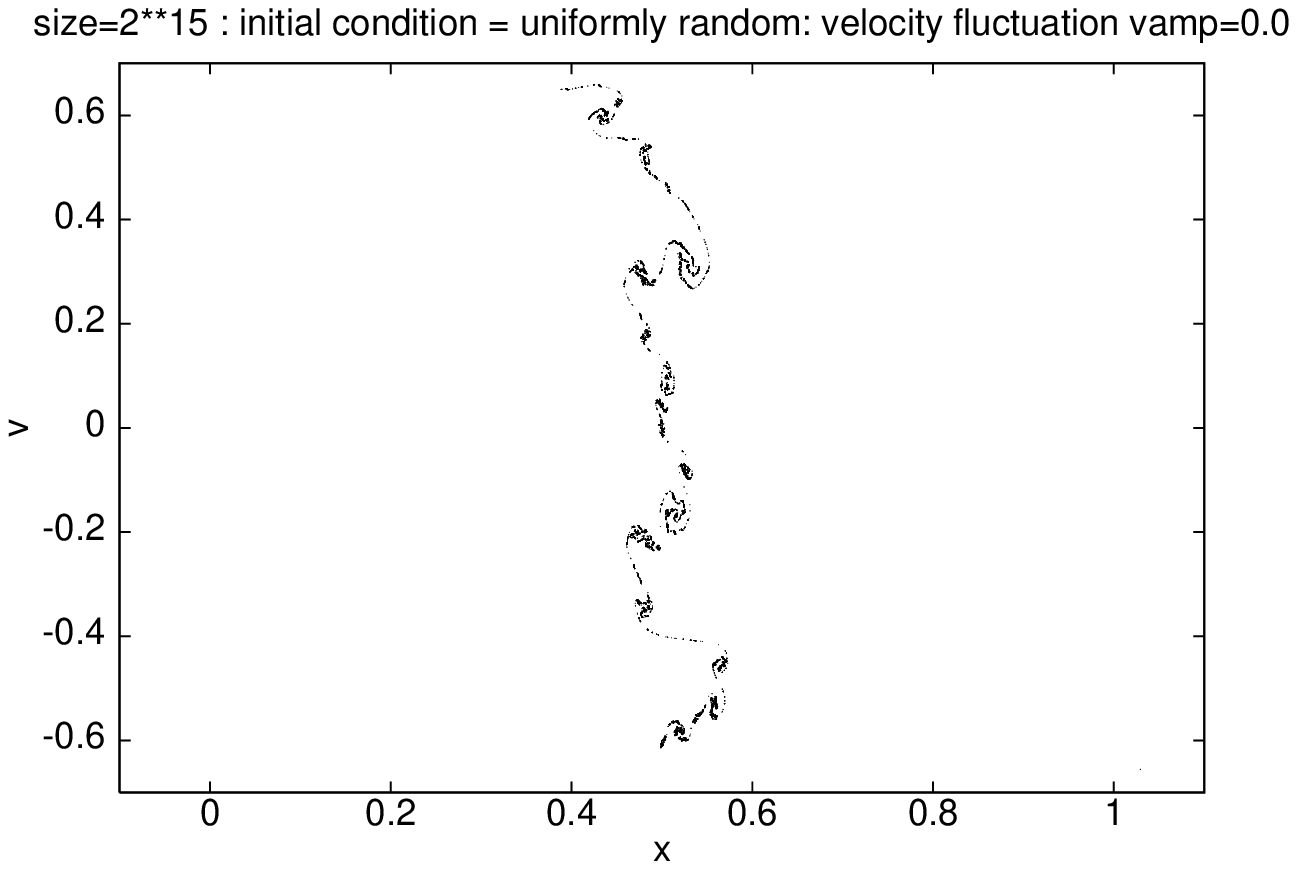}
    \includegraphics[width=7cm]{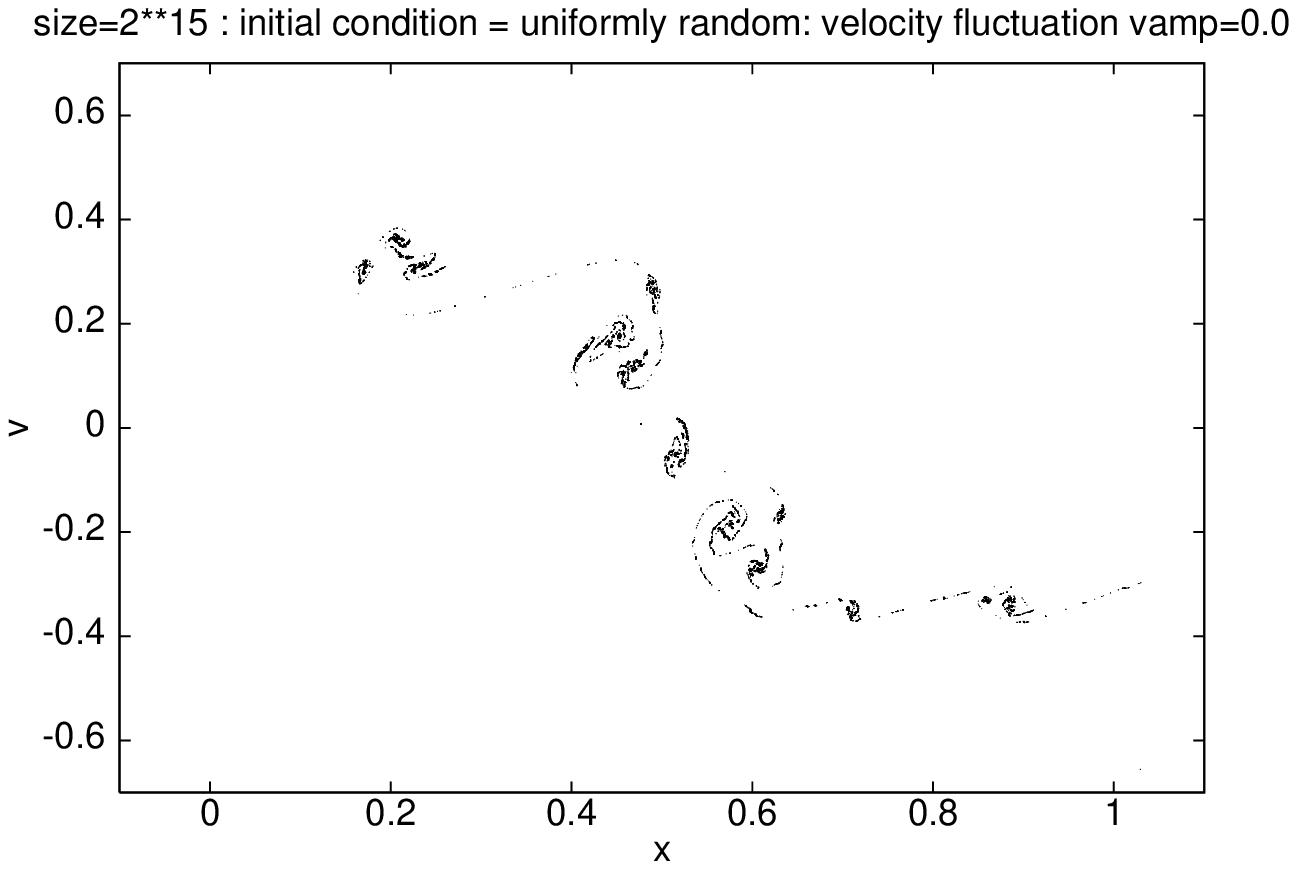}
    \caption{Formation of fractal structure. Initial condition :
      $x_i$ : uniformly random in [0,1], $u_i=0$ , $N=2^{15}$.
      Time are 2.34375,4.6875,7.03125, and 9.375. }
    \label{fig:xu-vamp=0}
  \end{center}
\end{figure}

\begin{figure}[hbtp]
  \begin{center}
    \includegraphics[width=7cm]{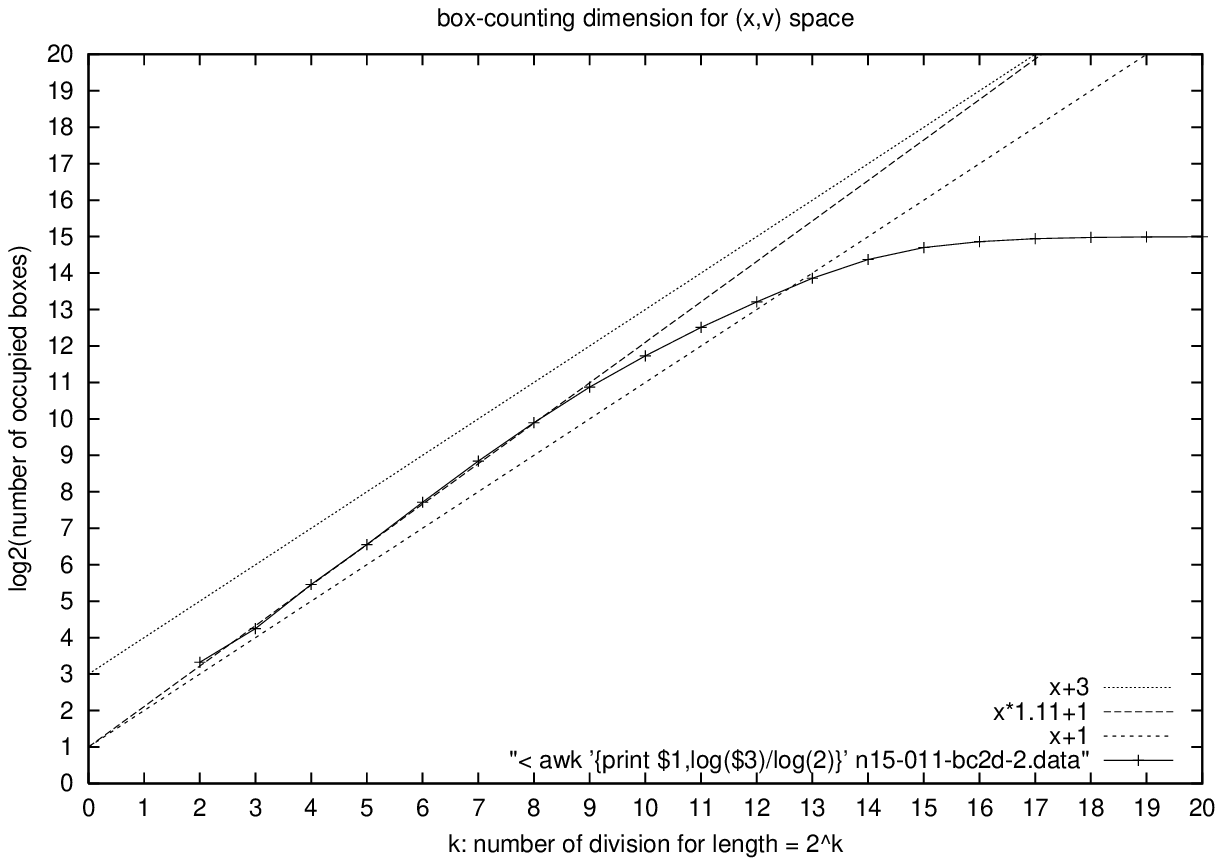}
    \caption{Box counting dimension of the $\mu$-space
      distribution shown in the bottom of Fig.\protect\ref{fig:xu-vamp=0}.
      Horizontal axis represents index of division $k$ of the phase space,
      i.e., 2-dimensional $\mu$-space is divided into $2^{2k}$
      of equal square cells 
      whose length $\ell$ of the side is proportional to $2^{-k}$. 
      Vertical axis represents $\log_2$ of 
      number $n(k)$ of the cells which contain
      at least one particle.
      Sample orbits  with the same class of initial condition with 
      different random number gives 
      dimension $D=1.1 \pm 0.04$ for 24 samples. Lines with $D=1$ are 
      also shown for comparison.}
    \label{fig:box-counting}
  \end{center}
\end{figure}

Fig.~\ref{fig:xu-vamp=0} show the evolution of the 
system~(\ref{eq:sheet-hamiltonian}) with $N=2^{15}$. 
Initial condition is chosen to be $V_r=0$ as
$x_i =\,\mbox{uniformly random}\, \in [0,1] $ 
and $u_i = 0$. 
That is, each sheet is placed randomly with zero velocity dispersion. 
The figures represent particle distribution
in $(x,u)$ space ($\mu$-space).  In the course of time evolution we see that
fractal structure is formed.
In Fig.\ref{fig:box-counting} we show box counting dimension of the
$\mu$-space distribution shown in the bottom of Fig.~\ref{fig:xu-vamp=0}.
Dimension is $D=1.1$.

\begin{figure}[hbtp]
  \begin{center}
 \includegraphics[width=6cm]{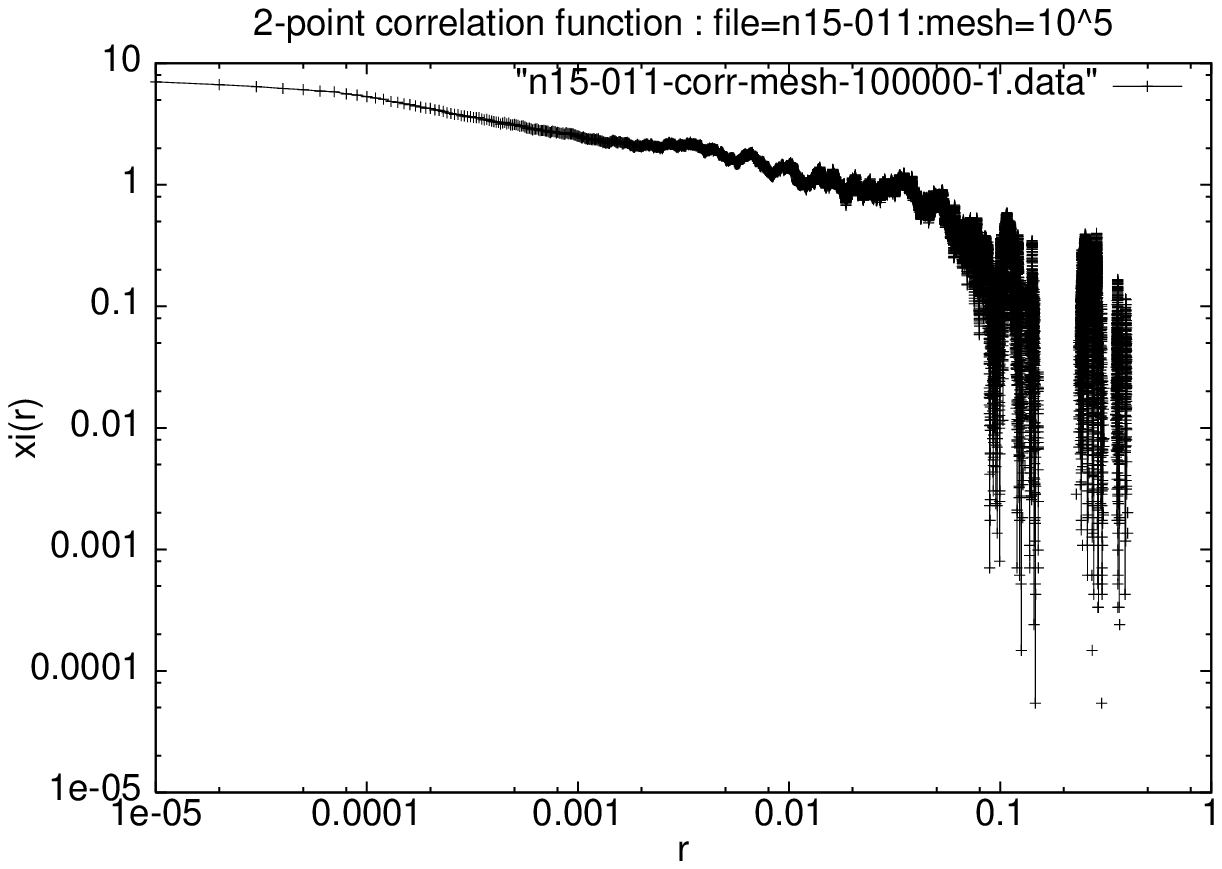}
\includegraphics[width=6cm]{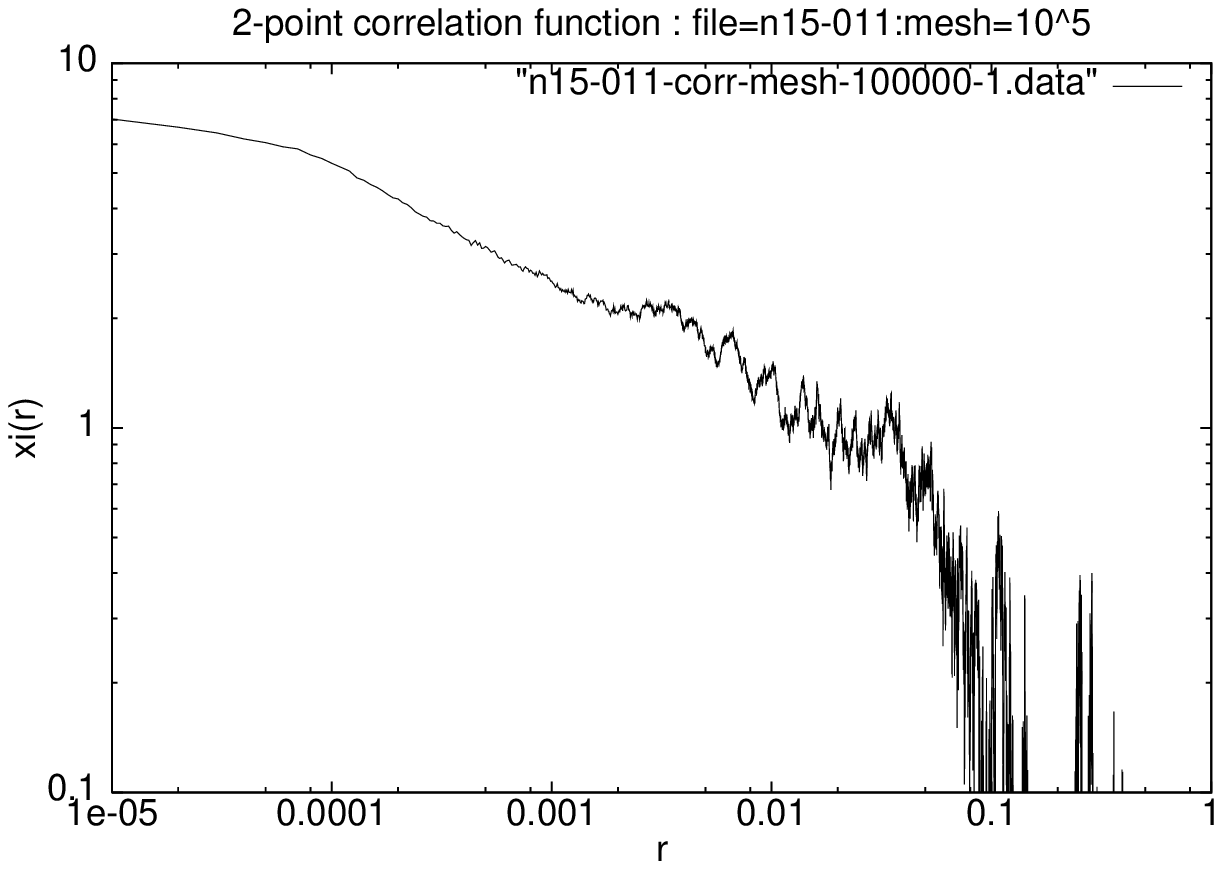}

    \caption{two-point correlation function $\xi(r)$ for $t=9.375$ in 
      Fig.~\protect\ref{fig:xu-vamp=0}. 
 These two figures are identical
       except for the range   of vertical axes.
      Exponent $\alpha$ of $\xi\propto r^{-\alpha}$ is
      $\alpha = 0.20 \pm 0.03 $ for 24 samples.
}
    \label{fig:corr-vamp=0}
  \end{center}
\end{figure}

Fig.~\ref{fig:corr-vamp=0} is a two-body correlation function $\xi(r)$ 
at $t=9.375$ in Fig.~\ref{fig:xu-vamp=0}. $\xi(r)$ is defined as
\begin{equation}
dP = n dV (1 + \xi(r))
  \label{eq:2-body-correlation}
\end{equation}
 where $dP$ is probability to find a sheet in  a volume $dV$ at
 distance $r$ apart from a sheet, and $n$ is the average number density.
We can see clear power-law behavior in $\xi(r)$ in 
Fig.~\ref{fig:corr-vamp=0}.

\begin{figure}[hbtp]
  \begin{center}
    \includegraphics[width=6.5cm]{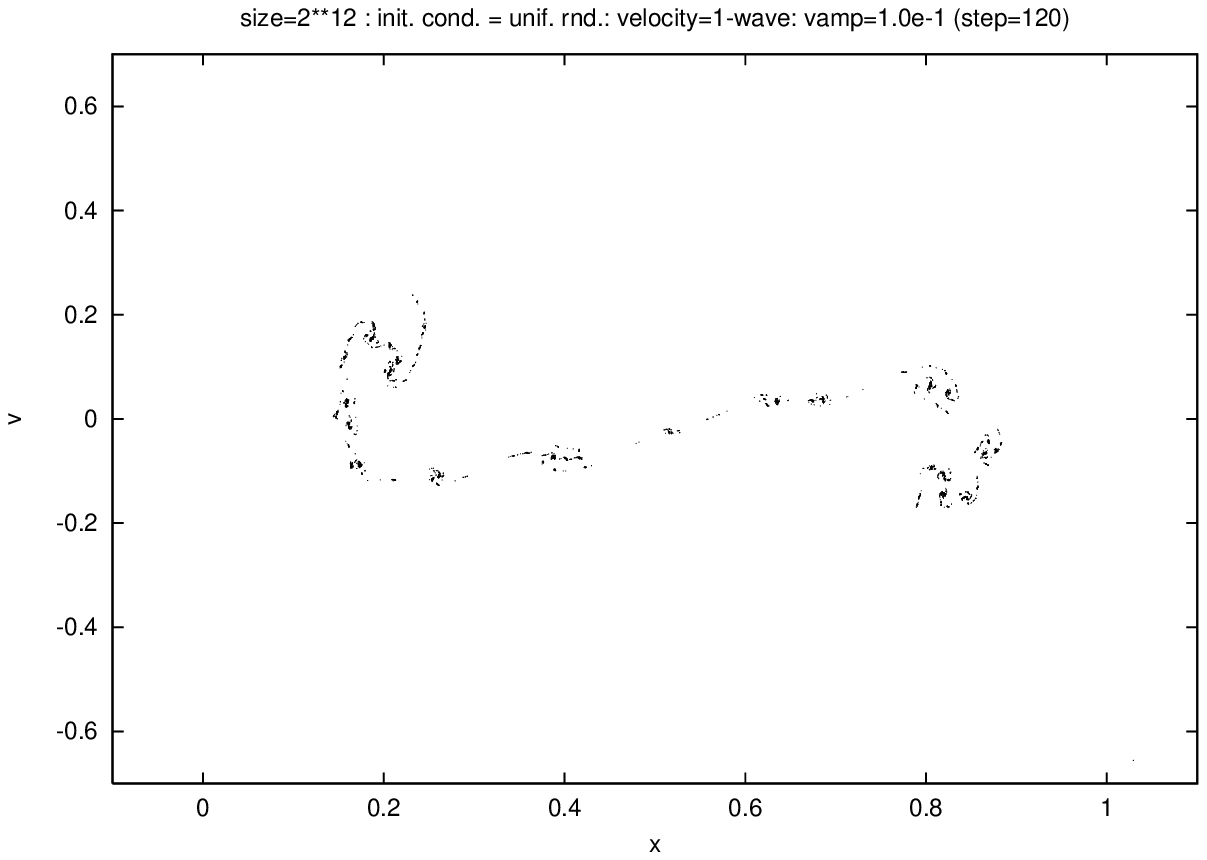}
    \includegraphics[width=6.5cm]{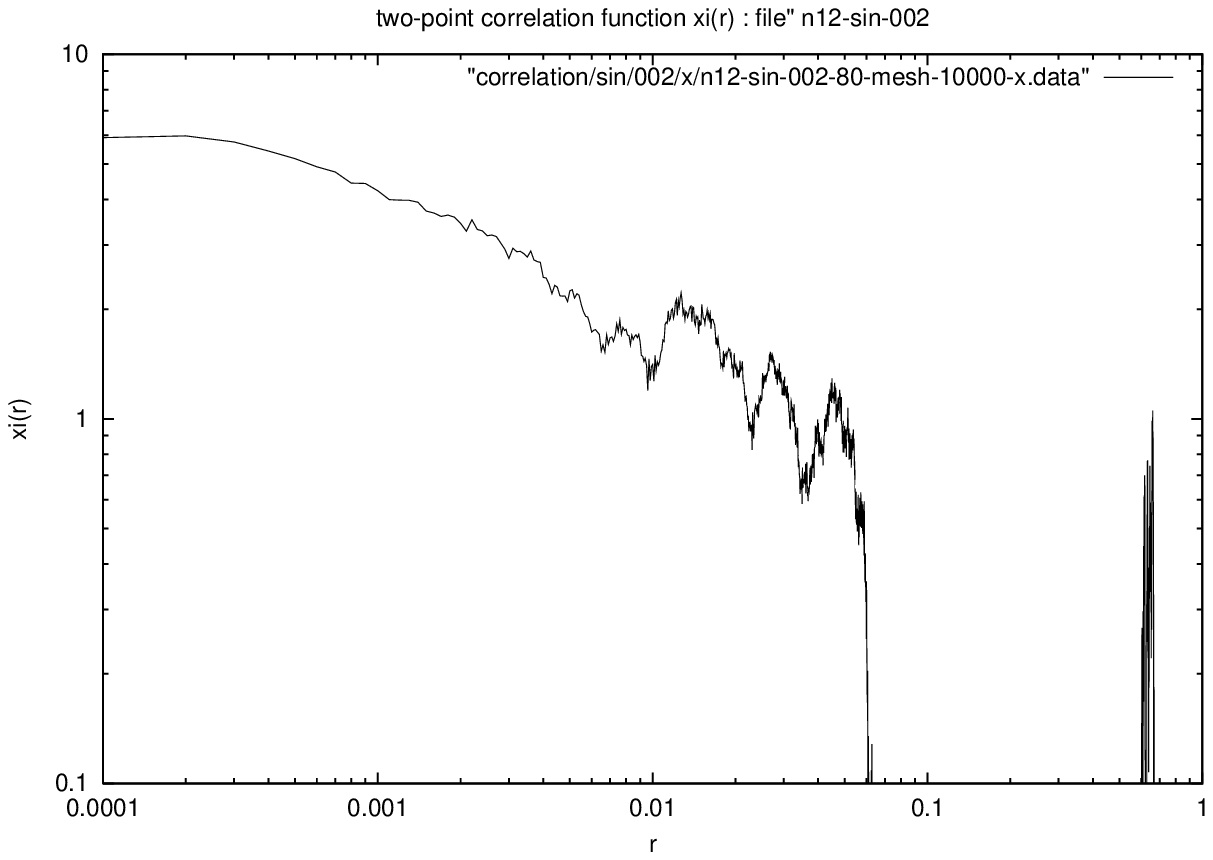}
    \caption{Distribution in $(x,u)$-space  (left) and correlation
function $\xi(r)$ (right)
for initial condition
      $x_i$ : uniformly random in [0,1], $u_i=0.1\sin2\pi (x_i-1/2)$ , 
      $N=2^{12}$. $t$=6.25. Boxcounting dimension is $D=0.95\pm 0.02$
      for 24 samples.
      Exponent $\alpha$ of the correlation function
      $\xi\propto r^{-\alpha}$ is
      $\alpha = 0.33 \pm 0.06 $ for 24 samples.
}
    \label{fig:xu-sin-002-080}
  \end{center}
\end{figure}

This power-law correlation
is also formed if we modify the initial condition
as
$x_i =\,\mbox{uniformly random}\, \in [0,1] $ 
and $u_i = a\sin2\pi (x_i-1/2)$, as shown in
Fig.~\ref{fig:xu-sin-002-080}.  In this initial condition
initial velocity of each particle is uniquely determined by the 
initial position of the particle, hence there is no velocity dispersion.

\begin{figure}[hbtp]
  \begin{center}
    \includegraphics[width=9cm]{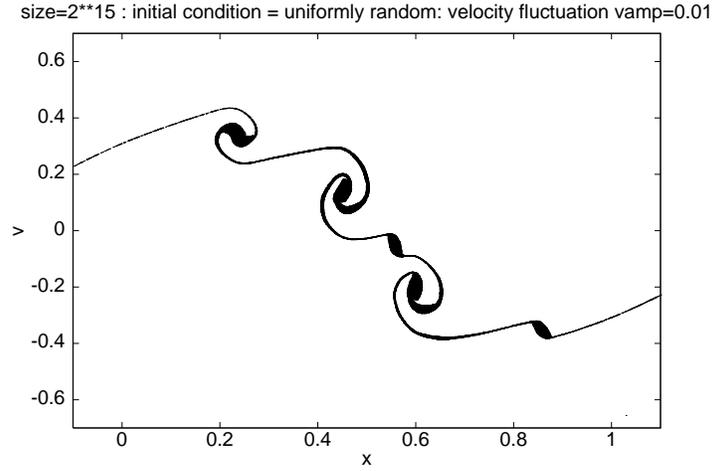}
    \caption{Example for $V_R \ne 0$. Fractal structure is not 
      formed. 
      Initial condition :
      $x_i$ : uniformly random in [0,1], $u_i=
      \mbox{random} \in [-0.01,0.01]$ ,
      $N=2^{15}$.
      Time = 9.375. }
    \label{fig:xu-vamp=0.01}
  \end{center}
\end{figure}

If we slightly increase the virial ratio from 0, the 
power-law correlation is not created. Fig.~\ref{fig:xu-vamp=0.01} show the 
$\mu$-space for an initial condition 
$x_i =\,\mbox{uniformly random}\, \in [0,1] $
and $u_i = \, \mbox{uniformly random}\, \in  [-0.01,0.01]$.
Here we see that fine structure observed in Fig.~\ref{fig:xu-vamp=0}
is no longer  observed. Flatness of two-point correlation function 
$\xi(r)$ in Fig.~\ref{fig:corr-vamp=0.01} supports the impression that
structure here is not  fractal but just made of  lumps with  
constant density. This type of state is also observed 
in \cite{hohl-feix-1967} and \cite{yamashiro-gouda-sakagami}. 

\begin{figure}[hbtp]
  \begin{center}
 \includegraphics[width=6cm]{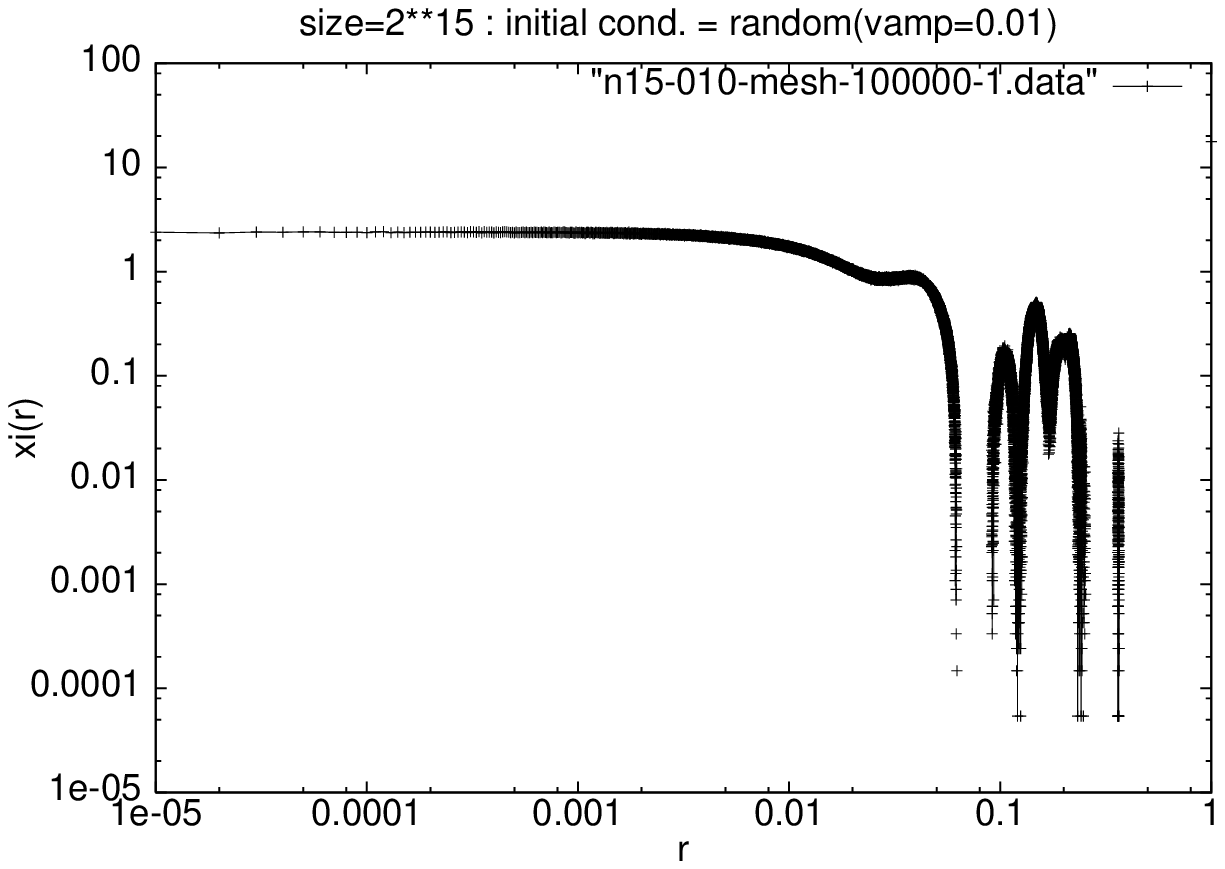}
\includegraphics[width=6cm]{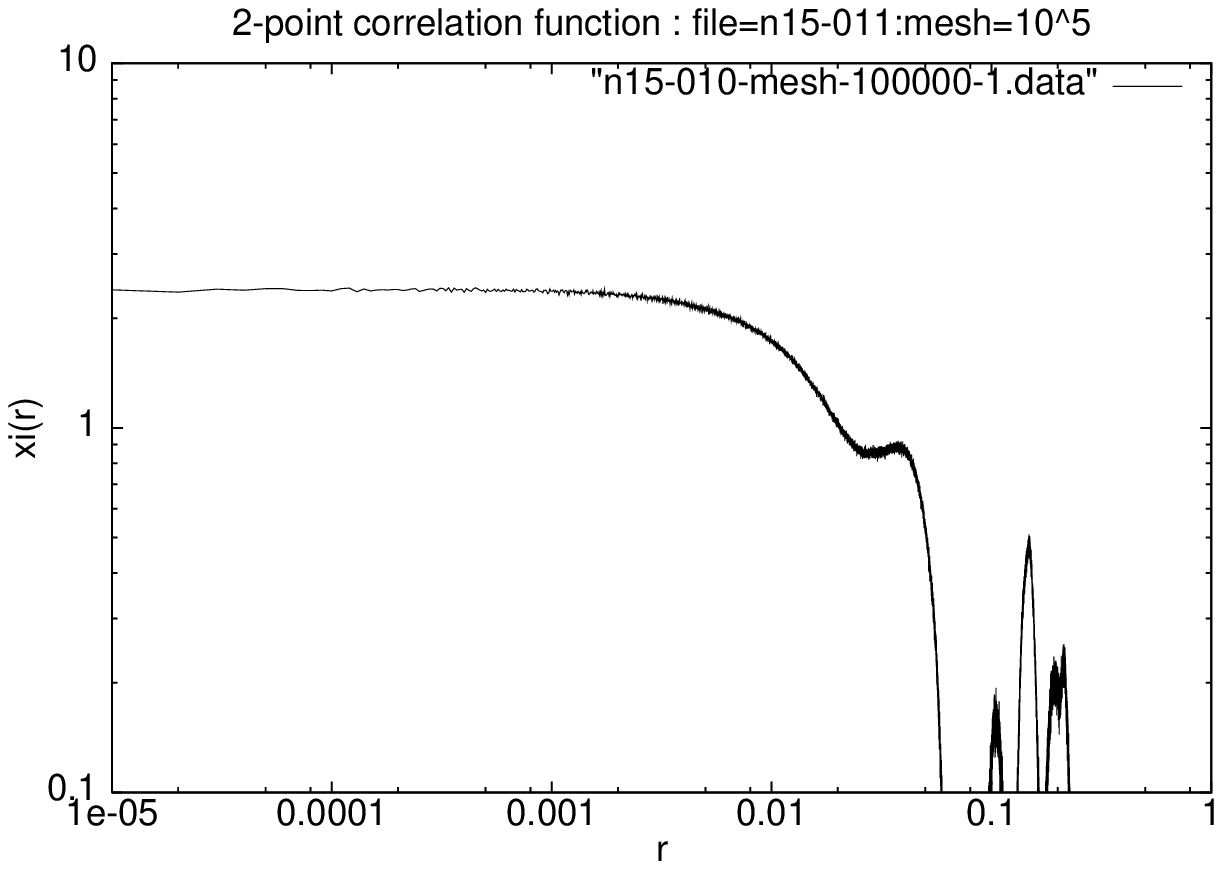}

    \caption{two-point correlation function $\xi(r)$ for $t=9.375$ in 
      Fig.~\protect\ref{fig:xu-vamp=0.01}. 
 These two figures are identical
      except for the range of   vertical axes.
}
    \label{fig:corr-vamp=0.01}
  \end{center}
\end{figure}

The fractal structure seen in Fig.~\ref{fig:xu-vamp=0} is not
a stationary state. As shown in Fig.~\ref{fig:xu-vamp=0-long},
the fine structure once created is gradually dissolved into 
large clusters. The reason for the destruction of this structure
is not clear now.

\begin{figure}[hbtp]
  \begin{center}
    \includegraphics[width=7cm]{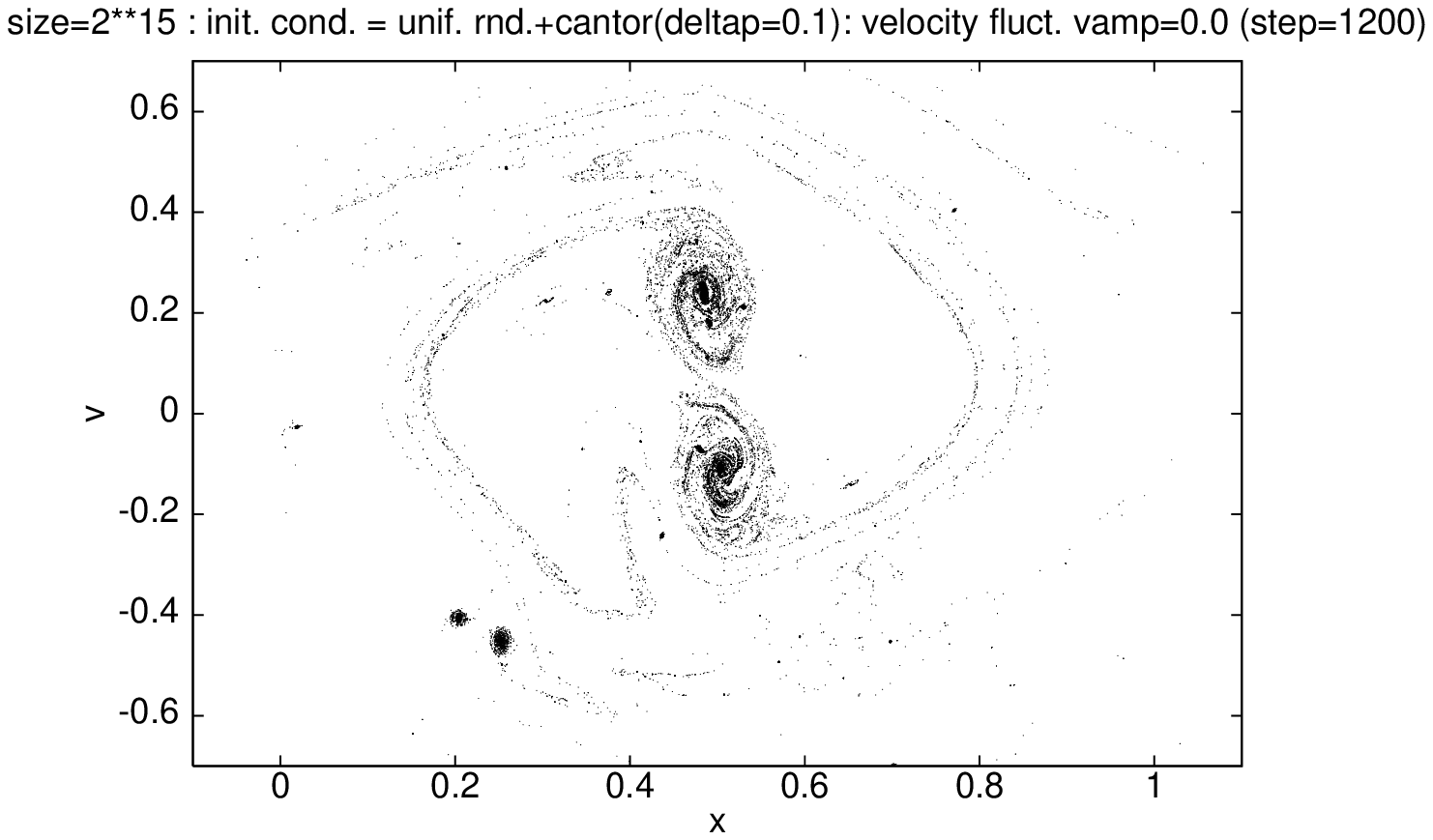}
    \caption{Continued from Fig.~\protect\ref{fig:xu-vamp=0}.
      Time=93.75.}
    \label{fig:xu-vamp=0-long}
  \end{center}
\end{figure}

\section{Summary and discussions}

In this letter we have shown that there is a new phase in 
the evolution of one-dimensional self-gravitating system 
where the system creates fractal spatial structure
from non-fractal initial conditions.
This novel phase appears for initial conditions with small virial ratio.
The structure is also formed for zero velocity dispersion
( i.e., `cold' condition).  Other initial conditions with large virial
ratio do not lead to  this phase.

As far as we observed, this phase is not a stationary state.
Whether or not this transiency is a real one or an artifact
caused by, for example, our computational 
limitation ( finiteness of number of particles and so on ) 
is not clear now.

Several studies revealed that two-point correlation function of young stars
in star-forming regions obey power-law~\cite{Larson,Hanawa-cluster}. 
The fractal structure we have 
found may be accounted for an  origin of the power-law.
Also it is a subject of future study if the phase
can be found in other  systems, such as 
galaxies in the course of evolution.

Similar structure is observed in a model with expanding 
universe~\cite{feix-1991-fractal,tatekawa-maeda-preprint}.  
It will be an interesting subject to examine the effect of 
expanding universe on the structure formation.

We would like to stress the importance of the fact that, 
to create fractal structure, it is not necessary for initial conditions
to be fractal.
Fractal structure does not have
characteristic spatial scale, nor does Newtonian potential. Hence 
it may not be so surprising that fractal structure can be found
in systems interacting with Newtonian potential.
However the relation between these two scale-free objects,
Newtonian potential and fractal structure, are still unknown. 
If the initial condition is fractal,
since the Newtonian interaction does not have characteristic 
length scale, the self-similar property of the initial condition
would be conserved. In fact, initial conditions with exact self similarity
(e.g., Cantor set) evolve in time while keeping the self-similar 
spatial structure.
On the other hand,
we have found that non-fractal initial conditions also 
leads to  fractal structure,  hence  in this case 
the structure is created by the dynamics itself.

The search for the mechanism why the fractal structure 
emerges  will be quite interesting.   The emergence is
interesting because it is an asymptote to scale-free behavior.
The emergence is also unique
in that this is a structure formation in conservative system 
against the tendency to thermal relaxation~\cite{tk-cluster,onion}.

We would like to thank Naoteru Gouda for valuable comments and 
discussions, and Yuji Masutomi for useful discussions.
We also thank Toshio Tsuchiya for advice on numerical methods.

\bibliographystyle{alpha}

\end{document}